\documentstyle[12pt]{article}
\begin{document}
\pagenumbering{roman}
\vspace*{.5in}
\begin{center}
\noindent
{\Large\bf  Semiflexible Chains under Tension}

\vspace{.25in}

{\it B.-Y. Ha and D. Thirumalai \\
Institute for Physical Science and Technology\\
University of Maryland\\
College Park, MD 20742}\\

\vspace{1cm}

{\bf ABSTRACT}
\end{center}
A functional integral formalism is used to derive the extension of a
stiff chain subject to an external force.  The force versus extension
curves are calculated using a meanfield approach in which the hard
constraint ${\bf u}^2(s)=1$ is replaced by a global constraint $\bigl<
{\bf u}^2(s) \bigr>=1$ where ${\bf u}(s)$ is the tangent vector
describing the chain and $s$ is the arc length.  The theory
{\it quantitatively} reproduces the experimental results for DNA that is
subject to a constant force.  

We also treat the problems of a semiflexible chain in a nematic
field.  In the limit of weak nematic field strength our treatment
reproduces the exact results for chain expansion parallel to the
director. When the strength of nematic field is large, a situation in
which there are two equivalent minima in the free energy, the
intrinsically meanfield approach yields incorrect results for the
dependence of the persistence length on the nematic field.  \\

\newpage
\pagenumbering{arabic}
\noindent
{\bf I.  Introduction} \\

Inspired by the elegant experiments by Smith et al.~\cite{SFB} on the
response of 
DNA to stetching by a constant force few theoretical papers [2-5] have
considered the effect of external field on semiflexible chains.  Since
DNA is stiff it can be described, at least approximately, as a worm
like chain.  The eariest theoretical papers dealing with this problem
were initiated by Fixman and Kovac~\cite{FK}.  These authors used a modified
version of the Gaussian model for stiff chains and provided
expressions for the stretching  as a function of applied force.
Their treatment is only valid when the applied force is small and significant
deviation from these predictions are observed at sufficiently large
values of the external force.  Marko and Siggia~\cite{MS} have calculated the
extension as a function of force for worm like chains and found that
their results fit the experimental data very well.  Some aspects of
this theory have also been considered by Odijk~\cite{Od} who also discusses the
competition between entropically dominated effects and elasticity
effects. 
 
In this paper we revisit the problem of a semiflexible chain subject to
tension.  A functional integral formalism together with a meanfield
treatment allows us to set up a general way of tackling problems
involving semiflexible chains.  When our theory is applied to the case
of stretching of DNA by a constant force, we obtain excellent
agreement with experimental results for the force versus extension. \\

\noindent
{\bf ${ {\bf I\!I}}$.  Elastic Response of a Semiflexible Chain}\\
\noindent
(a) {\it Meanfield Theory} 

The simplest model for a semiflexible chain (SC) is obtained by taking
into account the persistence in the tangential direction.  A SC can be
described by a space curve ${\bf r}(s)$ of total contour length $L$
with $s$ being the arc length.  The unit tangent vector is ${\bf u}(s)
\equiv {\partial {\bf r}(s)/ \partial s}$ with the constraint
that ${\bf u}^2(s)=1$ for all $s$.  Since the molecule is stiff it
costs energy to bend and the bending energy per segment length is
proportional to $({\partial {\bf u}(s) \over \partial s})^2$.  The
difficulty with calculations involving this formulation of SC is that
one has to invoke the constraint that ${\bf u}^2(s)=1$.  This makes
even the free chain problem non-linear.  There has
been extensive treatment of models of SC in the literature [6-15].  

Now
consider applying an 
external field that stretches the chain.  The effective free energy can
be written as 
$$
F=-k_{\rm B}T \ {\rm ln} \ Z   
\eqno(1a) $$
where 
$$ Z= \int {\cal D}[{\bf u}(s)] \ \delta ({\bf u}^2(s)-1) \ {\rm
e}^{-{\cal H}/k_{\rm B}T}
\eqno(1b) $$
with
$${{\cal H} \over k_{\rm B}T}={l_{\rm p} \over 2} \int \Bigl( {\partial {\bf
u}(s) \over \partial s} \Bigr)^2 {\rm d}s-\int{\bf f}(s) \cdot {\bf u}(s) {\rm d}
s .\eqno(1c)
$$
In Eq.(1c) $l_{\rm p}$ is the persistence length of the semiflexible
chain (which in the experiments of Smith at al.~\cite{SFB} on DNA is estimated to
be 53 nm).  For generality we have assumed that the external field
${\bf f}(s)$ depends on the arc length $s$.  The case of uniform ${\bf
f}(s)={\bf f}$ is appropriate for the experiments of Smith et
al. and is the one treated in the previous theoretical papers.  In
particular Marko and Siggia~\cite{MS} have used an eigenfunction expansion to
obtain the extension versus $f$ by exploiting the analogy between
this problem and the quantum problem of a dipolar rotor in an external
electric field.

Here we use a meanfield approach~\cite{HT,OEV} that effectively replaces the local
constraint ${\bf u}^2(s)=1$ by a global one $\bigl< {\bf u}^2(s)
\bigr>=1$.  Such a theory has been shown to give an exact expression
for the mean end-to-end distance of SC when ${\bf f}(s)=0$~\cite{HT}.  The basic
idea is to enforce the constraint ${\bf u}^2(s)=1$ using an auxillary
field variable
$\lambda(s)$ and evaluate the resulting integrals over $\lambda(s)$ by
the
stationary phase approximation.   The free energy
(cf. Eq. (1)) can be rewritten as  
$$\quad {\rm exp}(-F/k_{\rm B}T)=\int_{- i \infty}^{i \infty}{\cal
D}[\lambda (s)]  \int {\cal D}[{\bf u}(s)] \ \Psi[{\bf u}(s),{\bf f}(s)]
 \ {\rm exp} \Bigl[ \int_0^L \lambda(s) {\rm d}s \Bigr] +{\rm const.}
$$
$$= \int_{-i \infty}^{i \infty} \lambda(s) {\rm d}s \ {\rm exp}\{-{\cal
F}[\lambda,{\bf f}(s)] \} +{\rm const.} \qquad
\eqno(2a)$$
where
$$\Psi[{\bf u}(s),{\bf f}(s)]={\rm exp} \Bigl[-\mbox{$\frac{1}{2}$}
\int_0^L \int_0^L {\bf u}(s) \ Q(s,s') \ {\bf u}(s') + \int_0^L {\bf
u}(s) \cdot {\bf f}(s) {\rm d}s \Bigl]
\eqno(2b)$$
and
$${\cal F}[\lambda(s),{\bf f}(s)]=-{\rm ln} \ \int {\cal D}[{\bf
u}(s)] \ \Psi[{\bf u}(s),{\bf f}(s)]-\int_0^L \lambda(s) {\rm d}s 
\eqno(2c)$$
with
$$
Q(s,s')=\Bigl[ -l_{\rm p} \Bigl({\partial \over \partial s} \Bigr)^2 +
2 \lambda(s) \Bigr] \delta(s'-s)
.\eqno(2d)$$
The constants in Eq. (2a) comes from with normalizations associated
with ${\cal D}[\lambda(s)]$ and ${\cal D}[{\bf u}(s)]$.  These constants
will be ommitted from now on. 

The function ${\cal F}[\lambda(s),{\bf f}(s)]$ is a generating
functional which can be used to calculate various correlation
functions.  For example the connected correlation function
$$ \qquad \bigl< {\bf u}(s) \cdot {\bf u}(s') \big>_c=\bigl< {\bf u}(s) \cdot
{\bf u}(s') \bigr>-\bigl< {\bf u}(s) \bigr> \cdot \bigl< {\bf u}(s')
\bigr>
\eqno(3a)$$
$$\qquad \equiv -{\partial \over \partial {\bf f}(s)} \cdot {\partial {\cal
F} \over \partial {\bf f}(s')}
\eqno(3b)$$
$$=
3 Q^{-1}(s,s')
.\eqno(3c)$$
Similarly $\Delta R^2 = \bigl< R^2 \bigr>-\bigl< R \bigr>^2$ can be
written as 
$$\Delta R^2= \int_0^L {\rm d}s \int_0^L {\rm d}s' {\partial \over
\partial {\bf f}(s)} \cdot {\partial {\cal F} \over \partial {\bf
f}(s')}
.\eqno(4)$$
By performing the functional integrations with respect to ${\bf u}(s)$
the free energy ${\cal F}[\lambda(s),{\bf f}(s)]$ can be written as 
$${\cal F}[\lambda(s),{\bf u}(s)]=\mbox{$\frac{3}{2}$} \ {\rm tr} \
{\rm ln} Q - \mbox{$\frac{1}{2}$} \int_0^L {\rm d}s \int_0^L {\rm
d}s' \ {\bf f}(s) \ Q^{-1}(s,s') \ {\bf f}(s')- \int_0^L \lambda(s) {\rm d}s
.\eqno(5)$$
The variable $\lambda(s)$, that enforces the constraint ${\bf
u}^2(s)=1$, is evaluated by stationary phase approach~\cite{HT,OEV}.  The stationary
phase condition is obtained by requiring that
${\partial {\cal F} \over \partial \lambda(s)}$ be an extremum.  This leads to the
equation
$$ \mbox{$\frac{3}{2}$}
\biggl({2 \over Q} \biggr)_{s,s} + \int_0^L {\rm d}s \int_0^L {\rm
d}s' \ {\bf f}(s') 
 \ Q^{-1}(s,s') \ Q^{-1}(s,s'') \ {\bf f}(s'')=1
.\eqno(6)$$
The value of $\lambda(s)$ that makes ${\cal F}[\lambda,{\bf f}(s)]$
stationary depends on the precise form of ${\bf f}(s)$.  We now
specialize to the condition that ${\bf f}(s)={\bf f}={\rm const}$, 
independent of $s$.  This is the situation that has been treated in
the literature and is assumed to be relevant to the experiments of Smith
et al.~\cite{SFB}.  For constant ${\bf f}$ the stationarity condition (Eq. (6))
gives a uniform value for $\lambda$.  More precisely Eq. (6) reduces
to 
$$\mbox{$\frac{3}{2}$} \Biggl( {1 \over -{l_{\rm p} \over
2}\Bigl({\partial \over \partial s' } \Bigr)^2 +\lambda}
\Biggr)_{s,s'} + {f^2 \over 4 \lambda^2}=1
.\eqno(7) $$
In order to evaluate the first term in Eq. (7) in a transparant manner
we use the boundary condition ${\bf u}(0)={\bf u}(L)$ and ${ \partial
{\bf u}(0) \over \partial s}= {\partial {\bf u}(L) \over \partial
s}$.  These conditions were implicitly assumed in Eq. (2b).  The
easiest way to compute the first term in Eq. (7) is in terms of an
eigenfunction expansion.  Let $\{|s \big> \}$ denote the eigenstate
with $s$ the curviliniear space label and let $\{ |n \big> \}$ be the
states that are Fourier conjugate to $\{ | s \big> \}$ in such a way
$${\partial |n \big> \over \partial s} = i \Bigl({ 2 \pi n \over L}
\Bigr) |n \big>
.\eqno(8)$$
The states $|n \big>$ are eigenstate of `` momentum'' such that $ \bigl<n
| s \bigr>= {1 \over \sqrt{L}} {\rm exp}({i 2 \pi ns \over L})$.  In terms of
these eigenstates the first term in Eq. (7) becomes
$$\mbox{$\frac{3}{2}$} \Biggl( {1 \over 
-{l_{\rm p} \over 2} \Bigl({\partial \over \partial s' } \Bigr)^2 
+\lambda}
\Biggr)_{s,s'}=\mbox{$\frac{3}{2}$} \big< s | 
\Biggl( {1 \over -{l_{\rm p} \over
2} \Bigl({\partial \over \partial s' } \Bigr)^2 +\lambda}
\Biggr) |s \big>  \qquad \qquad $$
$$\qquad \qquad \qquad =\mbox{$\frac{3}{2}$} \Biggl( {1 \over -{l_{\rm p} \over
2} \Bigl({ \partial \over \partial s' } \Bigr)^2 +\lambda}
\Biggr) |n \big> \big< n | s \big> $$
$$\qquad \qquad \qquad \quad=\mbox{$\frac{3}{2}$} \sum_{n= -\infty}^{\infty} 
 \Biggl( {1 \over \lambda L+
{l_{\rm p} \over 2} {(2 \pi n)^2 \over L } 
 }
\Biggr)
.\eqno(9)$$
If we use the identity
$$\sum_{n=1}^{\infty} { {\rm cos} \ nx \over n^2 + \alpha^2}={\pi \over
2 \alpha} \cdot { {\rm cosh} \ \alpha (\pi- x) \over {\rm sinh} \ \alpha
\pi}-{1 \over 2 \alpha^2} \eqno(10a)$$
the stationarity condition for $\lambda$ becomes
$$\mbox{$\frac{3}{4}$} \sqrt{2 \over l_{\rm p} \lambda}
\ {\rm coth} (\mbox{$\frac{1}{2}$}\Omega L )+{ f^2 \over 4
\lambda^2}=1
\eqno(10b)$$
with $\Omega= \sqrt{2 \lambda \over l_{\rm p}}$.  For large $L$
Eq. (10b) simplifies to 
$$ 1- \mbox{$\frac{3}{4}$} \sqrt{{2 \over l_{\rm p} \lambda}}={f^2 \over
4 \lambda^2}
.\eqno(10c)$$
As $f \rightarrow 0$ this stationarity condition coincides with the one derived
previously. \\

\noindent
(b) {\it Correlation Function and Mean Square Internal Distance}

 In terms of the stationarity solution of $\lambda$, the
various correlation functions can be computed.  For example
$$\bigl< {\bf u}(s) \cdot {\bf u}(s') \bigr> =3 Q^{-1}(s,s') +{f^2
\over 4 \lambda^2} \qquad \qquad $$
$$\qquad \qquad \qquad \qquad \qquad \qquad 
=\mbox{$\frac{3}{4}$} \sqrt{2 \over l_{\rm p} \lambda} \  { {\rm
cosh}[(L-2|s'-s|) \Omega /2] \over {\rm sinh}(\Omega L/2) }+{f^2 \over
4 \lambda^2}
.\eqno(11)$$
For $s=s'$ the above equation reduces exactly to the stationarity
condition for $\lambda$ (cf Eq. (10b)) leading to $\bigl< {\bf u}^2(s)
\bigr>=1$.  Thus our approach ensures that the constraint ${\bf
u}^2(s)=1$ is satisfied globally.  For $L \rightarrow \infty$ Eq. (11)
becomes 
$$\bigl< {\bf u}(s) \cdot {\bf u}(s') \bigr> = \mbox{$\frac{3}{4}$}
\sqrt{1 \over l_{\rm p} \lambda} \ {\rm e}^{-|s'-s| \Omega} +{f^2 \over
4 \lambda}
.\eqno(12)$$
The mean squared internal distance of the semiflexible chain under
tension can be obtained as 
$$ \bigl< {| {\bf r}(s')-{\bf r}(s) |}^2 \bigr> = \int_s^{s'} {\rm d}s_1
 \int_s^{s'} {\rm d}s_2 \bigl< {\bf u}(s_1) \cdot {\bf u}(s_2) \bigr>
.\eqno(13) $$
Substituting Eq. (12) in Eq. (13) yields
$$
\bigl< |{\bf r}(s')-{\bf r}(s)|^2 \bigr>=\int_s^{s'}\!\int_s^{s'}
 \bigl< {\bf u}(s_1) \cdot {\bf u}(s_2) \bigr>{\rm d}s_1 {\rm d}s_2
\nonumber \qquad \qquad \qquad \qquad \quad \quad $$
$$
\qquad \qquad \qquad \qquad \qquad \qquad=\mbox{$\frac{3}{4}$} \sqrt{{2 \over l_{\rm
p} \lambda}} 
\Bigl[{2 \over \Omega}|s'-s|-{2 \over \Omega^2 }
(1-{\rm e}^{-\Omega |s'-s|} ) \Bigr] \nonumber
+{ h^2 \over 4 \lambda^2}(s'-s)^2
.\eqno(14) $$
The above equation naturally suggests a length scale, $l_f$, below
which entropy factors dominates and above which the mechanical energy
associated with the orienting field dominates.  For small values of
the contour length $l=|s'-s|$ the field dependent term becomes
negligible whereas for long $l$ the third term in Eq. (14) dominates.
A scale $l_f$ is obtained by balancing the first and last terms of
Eq. (14) and is given by
$$
l_f = {3 \lambda(f) \over f^2}
.\eqno(15) $$
When $f l_f \ll 1$ then the stationary phase condition can be easily
solved and one gets ${\lambda(f) \sim 1/l_{\rm p}}$ and consequently
Eq. (15) becomes $l_f \sim 1/f^2 l_{\rm p}$.  Under these conditions
one can think of the semiflexible chain to break up into a sequence of
``blobs'' of effective segment length given by
$$
\xi_f \sim f^{-1}
.\eqno(16)
$$
The above result coincides with the blob size predicted by flexible
chain under tension.  The above blob length is called {\it tensile
screening length}~\cite{Pi}.  For Eq. (16) to be valid it is necessary that
within the length $\xi_f$ there should be a large number of segments
each of length $l_{\rm p}$.  Since the effect $f$ within $\xi_f$ is
negligible the chain on this length behaves effectively as a Gaussian
chain containing $l_f/l_{\rm p}$ units.  It is for this reason $\xi_f$
in the law $f$ limit coincides with the Pincus blob length.  In the
limit of $l_f f \rightarrow \infty$, Eq. (15) gives $l_f \sim 1/f$ and
hence the chain conformation is dictated by coupling to the
mechanical energy.  In this limit we expect the chain to be aligned
with the external field with almost complete suppression of
fluctuations.  The effect of $f$ is felt at all length scales, i.e.,
$l_f$ is effectively zero. \\

\noindent
(c) {\it Average Chain Extension}

The quantity of experimental interest is the average extension $z$ of
the chain parallel to the external field.  The average elongation $z$
is computed using 
$$ 
z=\Bigl< \int_0^L {\bf u}(s) \cdot {{\bf f} \over |{\bf f}|} \Bigr>
.\eqno(17)
$$
The statistical average in Eq. (17) can be conveniently expressed in
terms of the free energy functional ${\cal F}$.  When ${\bf f}$ is
uniform we get
$$ \qquad
z={ {\bf f} \over |{\bf f}|} \cdot { \delta {\cal F} \over \delta
{\bf f}}=f \int_0^L \Bigl[ -l_{\rm p} \Bigl({\delta \over
\delta s }\Bigr)^{2}+2 \lambda(f) \Bigr]^{-1} {\rm d}s $$
$$
=  {f L \over 2 \lambda (f)}
.\eqno(18)$$
In Eq. (18) we have shown the argument of $\lambda$ to emphasize the
dependence of $\lambda$ on $f$.  The above equation and the associated
stationarity condition (cf. Eq. (10c)) determines the average chain
extension.

The chain extension $z$ can be easily calculated for the case of
$l_{\rm p} f \rightarrow 0$ and for $l_{\rm p} f \rightarrow \infty$.
When $l_{\rm p} f \rightarrow 0$ we get
$$
z=f {R_0^2 \over 3}
\eqno(19)$$
where $R_0^2=2 l_0 L= 2 ({2 \over 3} l_{\rm p})L$ is the size of the
corresponding ideal chain with $l_0 \equiv {2 \over 3} l_{\rm p}$ the
meanfield persistence length~\cite{HT}.
The above result is also obtained for a 
Gaussian chain under tension.  The leading order correction to Eq. (19) can 
be obtained for $l_{\rm p} f \ll 1$ by expanding Eq. (18) in power of
$f$.  The stationarity condition for $\lambda(f)$ up to ${\cal
O}(f^2)$ becomes
$$
\lambda(f) \approx \lambda_0 + {f^2 \over 2 \lambda_0^2}
\eqno(20)
$$
where $\lambda_0$ is the stationary phase condition for $f=0$~\cite{HT}.  Thus
the average elongation becomes
$$
{z \over L} \approx f l_0 (1- \mbox{$\frac{8}{9}$} f^2 l_0^2)
.\eqno(21)$$
  In the small $f$ limit the entropy
considerations dominate the effects due to the orienting field.  

The
dependence of $z$ on $f$ is quite different in the opposite limit.  If
$f l_{\rm p} \sim 1$ then we expect the external field to be relevant
at all length scales.  This would suppress the chain fluctuations on
scale greater than $l_{\rm p}$.  When $l_{\rm p} f \rightarrow \infty$
the stationary phase condition for $\lambda$ has a solution
$\lambda^{-1} \sim 0$ resulting in $\lambda={1 \over 2}f $ (Cf. Eq. (10c)).
  In this limit $z/L \rightarrow  1$ and the chain assumes a rod
conformation.  In the limit of $l_{\rm p} f \gg 1$ we can
approximately solve Eq. (10c) to get
$$
\lambda(f) \approx \mbox{$\frac{1}{2}$} f \Bigl(1+\mbox{$\frac{3}{4}$}
\sqrt{1 \over l_{\rm p} f} \Bigr)
.\eqno(22)
$$
The chain extension in the limit of large $f$ becomes
$$
{z \over L} \approx  \Bigl(1-\mbox{$\frac{3}{4}$}
\sqrt{1 \over l_{\rm p} f} \Bigr)
.\eqno(23)$$
This result has already been noted in the literature~\cite{Od}.  The $f^{-1/2}$
behaviour for semiflexible chains is in contrast to the Gaussian case.
 Apart from a numercal factor, the result in Eq. (23), valid for
$l_{\rm p} f \gg 1$, coincides with that discussed recently by Odijk~\cite{Od}
for semiflexible chains near the rod limit.  The prefactor in front of
$1 / \sqrt{l_{\rm p} f}$ obtained by Odijk is ${1 \over 2}$ which is slightly
smaller than our result.  The reason may lie in the fact that our
treatment utilizes a stationary phase method to enforce the hard constraint
${\bf u}^2(s)=1$ globally.

In the limit of $l_{\rm p} f \gg1$ the chain fluctuations are
relatively small and can be expanded in terms of 
$\theta_f \equiv (\theta_x,\theta_y)$ where $\theta_f$ is the angle
between the tangent vector and the external field.  The average
extension is related to $\bigl< \theta_f^2 (s) \bigr>$ as
$${z \over L} \approx 1-\mbox{$\frac{1}{2}$} \bigl< \theta_f^2 (s) \bigr>
.\eqno(24)
$$
For the uniform field we expect $\bigl< \theta_f^2 (s) \bigr>$ to be
independent of $s$ hence $\bigl< \theta_f^2 (s) \bigr> \equiv
\bigl< \theta_f^2 \bigr>$.  Comparing Eq. (23) we get an estimation of
the mean fluctuations 
$$
\bigl< \theta_f^2 \bigr> =\mbox{$\frac{3}{2}$} \sqrt{1 \over l_{\rm p}
f}  \eqno(25)
$$
which also coincides with the result of Odijk~\cite{Od} apart from the numerical
factor.

In addition to  yielding the results reported in the literature our theory
offers a simple estimate of the extension $z/L$ as a function of $f$ for
arbitrary values of $f$.  This is achieved by similtaneously solving
Eq. (10c) and Eq. (18).  As a test of the utility of the stationary
phase approach we compare our theory with the experiments of Smith et
al.~\cite{SFB} on DNA.  Marko and Siggia~\cite{MS} have already shown that their
numerically exact force 
extension for the semiflexible chain yields excellent agreement with
experiments.  In Fig. (1) we plot $z$ versus $f$ using the parameters
appropriate for DNA.  For comparison a few points from the
experiments, as presented in Fig. (1) of Ref.~\cite{MS}, are also shown.  The
meanfield approach reproduces the data quantitatively.  \\

\noindent
{\bf ${\bf I\!I\!I}$. Conclusions} \\

In this article we have considered the problem of semiflexible chain
subject to an external field using a functional integral formalism.
The crux of our method hinges on replacing the local constraint that
${\bf u}^2(s)=1$ by a global constraint $\bigl< {\bf u}^2(s) \bigr>=1$
where ${\bf u}(s)$ is the tangent vector.  The intrinsically meanfield
approach has been shown to be successful in producing the
configurational properties of semiflexible chains in the absence of
the external field~\cite{HT,LNN,OEV}.  Here we have shown that the stationary phase
method of enforcing the global constraint yields excellent results
even when the chain is subject to tension.  The very good agreement
between the theory and the experiments on DNA subject to force is a
confirmation of this assertion.  

The approach we have described here is sytematic but is not without
limitations.  These limitations become evident by considering the
behaviour of a semiflexble chain in a nematic environment, a problem
that has received considerable attention in the literature [17-22].  A
meanfield Hamiltonian of a stiff test chain in a matrix of other
chains in a nematic state can be written as [19a]
$${\cal H}=\mbox{$\frac{1}{2}$} \int_0^L\!\int_0^L {\rm d}s {\rm d}s'{\bf
u}(s) Q(s,s') {\bf u}(s')- g \int_0^L u_z^2 (s) {\rm d}s
\eqno(26)
$$
where ${\bf u}(s)=(u_\perp (s),u_z(s))$ and $g$ (with the dimension of
length) is the strength of the 
nematic potential.  By following the procedure outlined in the
previous section the stationary equation (one that extremizes the
free energy with Eq. (26) as the Hamiltonian) becomes
$$
1=\mbox{$\frac{1}{2}$} \sqrt{2 \over l_{\rm p} \lambda}+\mbox{$\frac{1}{4}$} 
 \sqrt{2 \over l_{\rm p} (\lambda-g)}
.\eqno(27)
$$
The above equation ensures that  
$$
1=\bigl<{\bf u}_\perp^2(s) \bigr> + \bigl< u_z(s) \bigr> =\bigl< {\bf
u}^2(s) \bigr>
$$
instead of ${\bf u}^2(s)=1$ for all $s$.
The mean extention of the chain parallel and perpendicular
to the director axis can be calculated as 
$$
\bigl< R_\perp^2 \bigr>=\int_0^L\!\int_0^L \bigl< {\bf u}_\perp (s) \cdot
{\bf u}_\perp(s') \bigr> {\rm d}s {\rm d}s'
\eqno(28a) $$
$$\bigl< R_z^2 \bigr>=\int_0^L\!\int_0^L \bigl< { u}_z (s) \cdot
{ u}_z(s') \bigr> {\rm d}s {\rm d}s'
.\eqno(28b)$$
Consider the weak nematic limit, i.e., $g \rightarrow 0$.  For small
$g$ the stationary condition can be solved and the results can be used
to get $\bigl< R_\perp^2 \bigr>$ and  $\bigl< R_z^2 \bigr>$.  These
lead to 
$$
{\bigl< R_\perp^2 \bigr> \over \mbox{$\frac{2}{3}$} ( 2 l_0 L)} \approx
1-\mbox{$\frac{1}{3}$}g l_0 
\eqno(29a)
$$
$${\bigl< R_z^2 \bigr> \over \mbox{$\frac{1}{3}$}( 2 l_0 L)} \approx
1+\mbox{$\frac{1}{3}$}g l_0 
\eqno(29b)
$$
where $l_0=\mbox{$\frac{2}{3}$}l_{\rm p}$.  The persistence length
along the nematic field is increased by a factor of $(1+{1 \over 3} g
l_0)$ which compares well with the exact results (in the limit of $L
\rightarrow \infty $) of Warner et al.~\cite{WGB}.

A similar analysis for $g \rightarrow \infty$ suggests that chain
fluctuations perpendicular to the director are totally suppressed
whereas the persistence length along the director is predicted to increase
by a factor of 2.  This result is in contrast with the analysis of
several authors [17,19a,21] who have shown that the effective persistence length
grows exponentially as $(gl_{\rm p})^{1/2}$.  The nematic potential,
 $-g u_z^2$, has two deep minima at large $g$ at $u_z =\pm 1$.  Thus
in the large $g$ limit the chain configuration is dominated by
``tunneling `` between the two minima by instanton~\cite{KDN}.  Mathematically
the partition function is dominated by instanton contributions which
our meanfield theory fails to capture.   It is clear that if there is 
  symmetry breaking in the problem then the replacement of the hard
constraint by a global one can lead to incorrect results. \\

\vspace {0.1in}
\noindent
{\bf Acknowledgement}: 

This work was supported in part by a grant from the
National Science 
Foundation through grant number NSF CHE 96-29845. \\

\newpage

\begin{center}
{\bf Figure Caption}
\end{center}
\noindent
Fig.(1): The solid line is the force-extension curve obtained by
simultaneously solving Eq. (10c) and Eq. (18).  The parameters have
been chosen from the fit of force-extension curve for DNA reported by
Smith et al.~\cite{SFB}.  The value of $L=32.9 \mu {\rm m}$ and
$l_0=53{\rm nm}$.  The square
represents the experimental results of Smith et al. as reported in
Ref.~\cite{SFB,MS}.

\end{document}